\documentclass[10pt,twocolumn]{article}
\usepackage{amsmath}
\usepackage{amsfonts}
\usepackage{color}
\usepackage{graphicx}
\usepackage{epstopdf}
\usepackage{multirow}

\usepackage{color}
\usepackage{url}
\makeatletter
\newif\if@restonecol
\makeatother

\usepackage[ruled,vlined]{algorithm2e}

\newcommand{\system}[2]{\left\{ \begin{array}{#1} #2 \end{array}\right.}
\newcommand{\matrice}[2]{\left( \begin{array}{#1} #2 \end{array}\right)}
\def\Z{\mathbb{Z}}
\def\Q{\mathbb{Q}}
\def\one{\vec{1}}
\begin{document}

\title{Efficient Compilation to Event-Driven Task Programs}
\author{Beno{\^\i}t Meister, Muthu Baskaran, Beno{\^\i}t Pradelle \\
  Tom Henretty, Richard Lethin \\ Reservoir Labs \\ \url{lastname@reservoir.com}}

\maketitle
\begin{abstract}
As illustrated by the emergence of a class of new languages and runtimes, it is
expected that a large portion of the programs to run on extreme scale computers
will need to be written as graphs of event-driven tasks (EDTs).
EDT runtime systems, which schedule such collections of tasks, enable more
concurrency than traditional runtimes by reducing the amount of inter-task
synchronization, improving dynamic load balancing and making more operations
asynchronous.

We present an efficient technique to generate such task graphs from a
polyhedral representation of a program, both in terms of compilation time and
asymptotic execution time.
Task dependences become materialized in different forms, depending upon the
synchronization model available with the targeted runtime.

We explore the different ways of programming EDTs using each synchronization
model, and identify important sources of overhead associated with them. 
We evaluate these programming schemes according to the cost they entail in
terms of sequential start-up, in-flight task management, space used for
synchronization objects, and garbage collection of these objects. 

While our implementation and evaluation take place in a polyhedral compiler,
the presented overhead cost analysis is useful in the more general context of
automatic code generation.
\end{abstract}





\section{Problem and Context}

The race for hardware speed and low-power is bringing computers from embedded to
large scale into the ``Extreme scale'' era, in which high numbers of cores react
heterogeneously to their environment, and are constrained by their
global energy consumption. 
This imposes tall requirements on the software, which must be as parallel as
possible to take advantage of the cores, and also adaptable to changing core
capabilities and avoid wasting energy. 

One way to address this problem is to depart from the Bulk-Synchronous
Programming (BSP) model.
Ironically, while BSP has historically promoted parallelism by enabling simple
programming models such as loop parallelism and Single-Program Multiple Data
(SPMD) computations, the model now seems to stand in the way of the 
amounts of parallelism sought out.
First, bulk synchronizations (across iterations of a for loop, for instance) 
often express an over-approximation of the actual dependences among computation
instances (whether they are tasks or loop iterations).
Also, synchrony often results in a loss of parallelism and a waste of energy,
since cores spend a portion of their time waiting for some condition to happen
(e.g., a barrier to be reached by other cores, a spawned task to return). 

Thus, it is commonly believed that the next generation of parallel software will
be asynchronous and non-bulk. 
In other words, programs will be expressed as a graph of tasks, in which tasks
are sent for asynchronous execution (``scheduled''), and they become runnable
whenever their input data is ready.
In this model, the more accurate the inter-task dependences are with respect to
the semantics of the program, the more parallelism is exposed.

Some programming models support the expression of parallel programs as recursive
tasks, with for instance Cilk \cite{Blumofe:ACM:1995}, X10 \cite{x10}, and
Habanero \cite{habaneroj}. In these models, each task can only depend on one
(parent) task, or on the set of tasks scheduled by a sequential predecessor. 
A major advantage of these models is that they offer provable
non-deadlock guarantees.  
However, this comes at the cost of being less general than other systems which
express programs as acyclic graphs of tasks \cite{ocr-spec-100, swarm, CnC, kong14taco,
  legion, ompss, parsec}\footnote{Since dependences represent constraints on the
  task exectution order, a task graph needs to be acyclic for a valid execution
  order of its tasks to exist.}.
While these graphs exist in the literature under a variety of names, here we are
using the name ``Event-Driven Tasks'' (EDT) to refer to them. 
An event here represents the satisfaction of a dependence. 

Since with more generality and performance also come higher programming
difficulty, we have worked on tools to automatically {\em generate} such
programs, when they can be modeled with the polyhedral model. 
With this model, geared towards compute-intensive loop codes, the parallelization
tool is provided with a precise representation of the program, whose task graph
can be generated statically, as described in \cite{Baskaran2009,kong14taco}. 

Since dependences are determined statically at parallelization time, relying
on systems that discover dependences at runtime \cite{legion, ompss} would be
wasteful and is hence not considered in this paper.

One of the intrinsic challenges of automatic parallelization is to define
programmatic ways of generating tasks and dependences and of using the target
system capabilities without introducing too much overhead.
Another important challenge with polyhedral representations is to maintain
optimization time tractable, which requires the use of nimble operations on
polyhedra representing the tasks and their dependences.

This paper offers two main contributions related to the automatic generation of
EDT codes, and in particular the dependence relationships among tasks. 

After comparing run-time overheads implied by implementation strategies based
on a set of basic synchronization models available in current EDT runtimes,
we propose a nearly-optimal strategy based on a slight improvement of one of the
models, in section \ref{sec:deps-comparison}. 

Then, focusing on the case of the polyhedral model, we present a novel, scalable
technique for automatically generating tasks and dependences in section
\ref{sec:scalable}.

In section \ref{sec:deps-codegen}, we show how this model can be used to
generate EDT codes with the discussed synchronization models, along with further 
code optimizations. 
Finally, we evaluate the benefits of using our techniques on a set of
benchmarks in section \ref{sec:exp}, discuss related work in section
\ref{sec:related} and summarize our findings in section \ref{sec:conclusion}.

\section{A comparison of synchronization models} \label{sec:deps-comparison}


Throughout this paper, we care about the automatic, optimal generation of
EDT-based codes, and the cost of using various synchronization models. 
We are excluding other questions such as the per-task overhead of the
runtime, which boil down to the constraint of making the tasks large enough
(thousands to tens of thousands of operations per task seems to be the norm on
x86-based platforms).

Expressing a program as a graph of event-driven tasks (EDTs) requires some
amount of bookkeeping.
We are interested in overheads that such bookkeeping would entail, and in their
behavior as the number of tasks grows.
We illustrate these overheads by referring to the system proposed by
Baskaran et al \cite{Baskaran2009}. 
While we remind the reader that its authors did {\em not} intend it for a
large-scale system, it has been used in larger-scale works for automatic
parallelization to task graphs using the polyhedral model since then
\cite{dathathri14dataflow, kong14taco}. 

Let $n$ be the number of vertices in the task graph. 
One of the advantages of the reference method is its simplicity. 
A master thread sets up a graph of tasks linked by dependences; 
then, an on-line list scheduling algorithm defines when and where tasks get
executed. 

One obvious bookkeeping overhead in this scheme is that it requires a setup
phase {\em before} the program can actually run in parallel.  
Amdahl's law dictates that the cost of sequential part of bookkeeping
tasks must be insignificant as compared to the execution time of a task. 
The importance of this {\bf sequential start-up overhead} becomes greater as the
available parallelism grows, and is hence crucial to minimize.

Also, the spatial cost of representing the dependences has a major impact on
scalability, and even feasibility of generating executable programs. 
For instance, the baseline method represents inter-task dependences explicitly,
all at once, and hence has an $O(n^2)$ {\bf spatial overhead}.

Another source of overhead relates to the amount of tasks and dependencies the
runtime has to manage at a given point in time. 
EDT-based runtimes make it possible to run tasks asynchronously, i.e., a task
can {\em schedule} other tasks, even before its inputs are ready and without
waiting for its completion. 
It is the user's responsibility to let the runtime know when a task's inputs are
ready (i.e., when the task can be executed) through the synchronization constructs
provided by the runtime API. 
The amount of tasks that are scheduled before they are ready to execute
is referred to here as {\bf in-flight task overhead}.
The number of unresolved dependence objects that need to be managed by the
runtime at any time is the {\bf in-flight dependence overhead}. 

Finally, garbage collection of objects created for the purpose of running a
task graph may entail large overheads, especially if the garbage collection must
be done only after a large set of tasks has completed. 
Here, we measure {\bf garbage collection overhead} by the number of objects that
are not useful anymore but are not destroyed at any point in the execution.

In the next section, we go through the synchronization paradigms we have
experimented with, and examine the overheads induced by their use in automatic
code generation. 

\subsection{Synchronization constructs} \label{sec:sync-paradigms}
We have implemented task-graph code generation based on three synchronization
models that we observed in existing task-graph-based runtimes.
In one of them, a task (often called ``prescriber task'') sets up input
dependences for a task before these dependences can be satisfied by other
tasks. 
Here we call it the {\em prescribed} synchronization model. 

Alternatively, dependence satisfaction information can go through {\em tags},
objects that tasks can {\em get} from and {\em put} into a thread-safe table.
At a high level, tags are identified with inter-task dependences, and tasks
check that their input dependences are satisfied by {\em get}ting the
corresponding tag.
When a task satisfies another task's input dependence, it puts the corresponding
tag into the table. 
The structure of the tasks our tool generates using tags is defined by a
sequence of {\em get}s, then the task's computation, followed by a sequence of
{\em put}s.
The tag table mechanism is currently available in the SWift Adaptive Runtime
Machine~\cite{swarm} (SWARM) runtime.
It has also been proposed for future implementation in the Open Community
Runtime~\cite{ocr-spec-100} (OCR) 1.0.1 specification.

Finally, tasks may be associated with a counter, and scheduled upon the counter
reaching zero. 
In this case, only the number of dependences are represented, as opposed to a
set of dependences. 
A {\em counted dependence} is a synchronization construct that associates the
scheduling of a task with a counter. 
The task is scheduled whenever the counter reaches zero. 
We extended this construct to one that safely creates a counted dependence that
needs to be decremented if it does not exist yet, and called this construct
``autodec.''  

Table \ref{tab:sync-paradigms} presents a breakdown of the synchronization
constructs available in the Exascale-oriented runtimes studied in this paper. 
\begin{table} 
\begin{center}
\begin{tabular}{|c|c|c|c|}
\hline
{\bf Runtime} & Prescribed & Tags      & Counted    \\ \hline
OCR           & $\times$   &           & $\times$   \\ \hline
SWARM         &            & $\times$  & $\times$   \\ \hline
\end{tabular}
\caption{Task synchronization models, and examples of runtimes implementing
  them.} \label{tab:sync-paradigms}
\end{center}
\end{table}

In OpenStream, a main thread sets up ``streams'', which are both a
synchronization and communication queue between tasks. The order in which tasks
that write to a stream are scheduled by the main task defines which readers of
the stream depend on which writers. In that sense, the synchronization model here
is akin to a prescribed model. 
Tags are available in Intel's implementation of the CnC coordination
language \cite{CnC}, as well as prescription.
The \verb|async| and \verb|finish| constructs available in Cilk,
X10 and Habanero can be implemented using counted dependences.
Futures can be supported as well, for instance by adding prescription or tags. 

The next section studies the asymptotic overheads linked with the use of these
synchronization models. 

\subsection{Comparative overheads}
In this section, we go through four synchronization models and find out the
implied overheads when using them in automatic code generation. 
Different ways of using them are considered, when these lead to different
overheads. 
Our analysis is summarized in Table \ref{tab:overhead-summary}, where $n$ is the
number of tasks in the graph, and $r$ is the maximum number of tasks that are
ready to run simultaneously in any possible execution of the graph.
$d$ is the complexity of computing the number of predecessors to a task, 
and $o$ is the maximum number of dependences going out of a task (the maximum
out-degree in the task graph). 
\begin{table*}
\begin{center}
\begin{tabular}{|c|c|c|c|c|c|}
\hline
 ~      & Start-up & Spatial & In-flight tasks & In-flight deps & Garbage collection \\ \hline
Prescribed      & $O(n^2)$   & $O(n^2)$ & $O(n)$    & $O(n^2)$ & $O(n)$      \\ \hline
Tags Method 1   & $O(1)$     & $O(n^2)$ & $O(n)$    & $O(n^2)$ & $O(1)$      \\ \hline
Tags Method 2   & $O(1)$     & $O(n)$   & $O(n)$    & $O(n)$   & $O(n)$      \\ \hline
Counted         & $O(n.d)$   & $O(n)$   & $O(n)$    & $O(n)$   & $O(1)$      \\ \hline
Autodec w/o src & $O(1)$     & $O(n)$   & $O(n)$    & $O(r.o)$ & $O(1)$      \\ \hline
Autodec w/ src  & $O(1)$     & $O(r.o)$ & $O(r)$    & $O(r.o)$ & $O(1)$      \\ \hline
\end{tabular}
\end{center}
\caption{Overheads associated with task graph synchronization
  models}\label{tab:overhead-summary}
\end{table*}

The following sections (\ref{ssec:prescribed} through \ref{ssec:autodec}) go
through a detailed analysis for each synchronization model.
They ultimately show that optimal overheads can be obtained by using autodecs.

\subsubsection{Prescribed synchronization} \label{ssec:prescribed}
To support our discussion, let us consider a particular task in a task graph,
and let us refer to it as the {\em target task}.
We also define $n$ as the number of tasks in the graph to run on the 
runtime system.
For simplicity, we also assume a ``master'' thread/task/worker, which is
able to schedule tasks.
 
\begin{figure}
\begin{center}
\includegraphics[width=0.3\columnwidth]{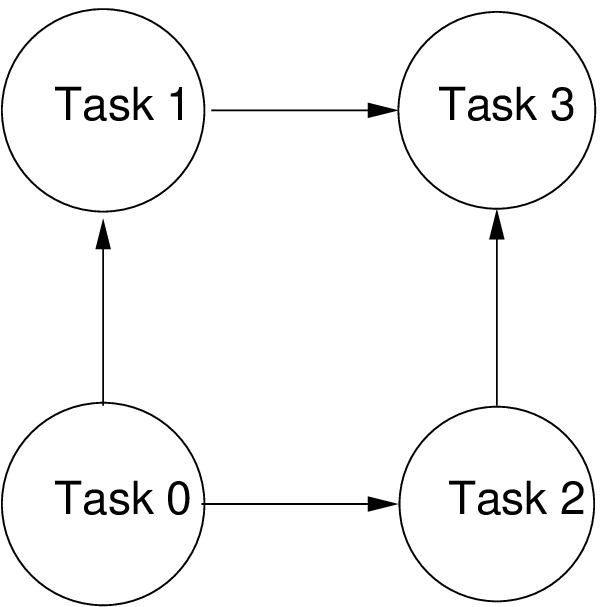}
\end{center}
\caption{Simple example where a successor task has two predecessors}\label{fig:diamond}
\end{figure}

With prescribed synchronization, the target task is created and its input
dependences are set up by a task that precedes it in the task graph. 
This method is straightforward for task graphs that are trees, but less so in
cases where tasks may have more than one direct predecessor. 
To illustrate this, consider the case when a task has more than one
predecessor, illustrated with a ``diamond'' pattern in Figure
\ref{fig:diamond}, and in which Task 3 is the target task. 
In this toy example, dependences do not define a particular order of execution
between Tasks 1 and 2. 
Task 3 needs to be created, and its input dependences set up by one of its
predecessors. 
Without further synchronization, it is impossible for Task 2 to know whether
Task 1 has created Task 3 already, or if it needs to be created, and conversely
for Task 1. 
When the target task has more than one predecessor, and without further
synchronization possibilities, there are three methods available: 
\begin{enumerate} 
\item[1.] A task that dominates all the transitive predecessors of the target task can
  be responsible for the creation of the target task. 
  The only dominator in our example is Task 0. 
  In our example, Task 0 would indeed be responsible for the setup of all the
  other tasks.   
\item[2.] Alternatively, one of the transitive predecessors to the target task is
  chosen to set up the target task, and additional dependences are introduced
  between the chosen task and the transitive predecessors to the target task
  that don't precede the chosen task.  
  The additional dependences materialize the fact that the created target task
  becomes an input dependence to the transitive predecessors. 
  In our example, we could for instance choose Task 1 as the creator and add a
  dependence between Tasks 1 and 2, making the task graph entirely sequential.
\item[3.] Finally, a special prescriber task can be added for the specific
  purpose of setting up the target task. 
  Dependences representing the created task object are added between the
  prescriber task and a set of tasks that transitively precede and
  collectively dominate the target task. 
  One instance of such set is the set of direct predecessors to the target
  task. 
\end{enumerate}

Since one of the fundamental goals of EDT runtimes is to increase the
amount of available parallelism, the loss of parallelism induced by the
introduction of sequentializing dependences~\footnote{
  ``sequentializing dependences'' are dependences that are not required by the
  semantics of the program and that reduce parallelism.} 
in Method 2 excludes it from the set of acceptable methods.

The worst case for Method 1 is when the task graph is dominated by one task, as
in the diamond example.
This occurs fairly frequently, including in many stencil computations as
parallelized through polyhedral techniques.
In this case, the dominator task is responsible for setting up all tasks in the
graph and their dependences, before any other task can start. 
An equivalent case is when all tasks are dominated by a set of tasks, in which
case the host is the only common dominator of all tasks. 
Both cases result in a $O(n^2)$ sequential overhead, which accounts for the
setting up of all the dependences in the graph. 

A naive implementation of Method 3 would generate a prescriber task for each
target task in the graph that has more than one predecessor, and introduce
dependences between the prescriber task and a dominating set of predecessors to
the target task. 
Notice that this process adds an input dependence to the predecessors, which
may have only had one predecessor before adding the prescriber task. 
These predecessor tasks now fall into the original problem, and themselves
require a prescriber task, which must precede the initial prescriber.
This results in a transitive construction of prescriber tasks.

\begin{figure}
\begin{center}
\includegraphics[width=\columnwidth]{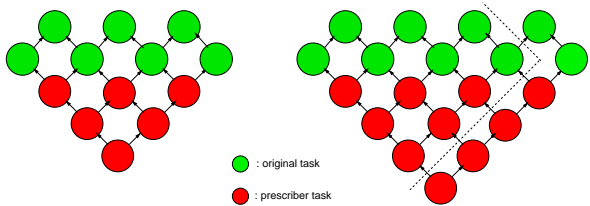}
\end{center}
\caption{Growth of prescriber tasks as introduced by Method 3.} \label{fig:meth3-scale}
\end{figure}

In the worst case, the number of such tasks grows as a polynomial of $n$,
$pr(n)$. 
This is illustrated in Figure \ref{fig:meth3-scale}, where the number of
prescriber tasks grows in $O(n^2)$. 
The number of dependences for these prescriber tasks is then expected to be in
$O(pr(n)^2)$. 
 
In the same example, while some tasks may start before all the prescribing tasks
have completed, the number of prescriber tasks that need to complete
before any single non-prescriber task in the graph equals $\frac{n-1}{2}$, i.e.,
a $O(n)$ sequential start-up overhead.
The complexity of the sequential start-up overhead appears to be a polynomial
of degree less than $pr(n)$. 

Such an approach would be impractical at large scale, not only because the
potentially high complexity of its incurred sequential overhead, but also
because many additional tasks may need to be created. 

A response to both these issues is to group the prescriber tasks into
``macro-''' prescriber tasks, which are responsible for setting up several
tasks. 
Without any particular knowledge of the graph structure, the optimal grouping
for these tasks is when creating one single prescriber task that sets up all the
graph. 
This solution is exactly equivalent to Method 1, and it has the same overheads. 

Analyzing the graph structure in the hope of finding more than one group with
lower overheads also has a $O(n)$ overhead. 
Hence Method 3 is at best as good as Method 1. 

We implemented Method 1 using OCR.
Its spatial overhead is $O(n^2)$ because all dependences are represented explicitly.
The number of tasks to be handled by the scheduler at once (in-flight task
overhead) is $O(n)$, the number of in-flight dependences is $O(n^2)$ and the
input dependence objects for each task can be garbage-collected when the task
starts.

\subsubsection{Tag-based synchronization} \label{ssec:tags}
Notionally, a tag is a key in an associative table.
A predecessor and its successor tasks synchronize through a tag that
represents the completion of the predecessor task to the successor task. 
The predecessor signifies its completion by putting the tag in the table. 
The successor can wait for tags to be available in the table by getting the
tag. 
Control returns to the task whenever all the tags for which a \verb|get| was issued
have been put in the table. 
To avoid deadlocks, \verb|get|s are typically asynchronous.
When a tag is put, all the tasks that did \verb|get| the tag are considered for
execution. 

We have found two meaningful tag-based methods to perform inter-task
synchronization.
\begin{enumerate}
\item [1.] In the first method, each pair of tasks linked by a dependence is
  mapped to a tag. 
\item [2.] In the second method, independently developed in
  \cite{vasilache13tale}, one tag is associated with each predecessor task.
  Before completion, each task puts a tag that signifies that it has completed.
  Its successors all get the same tag from the table.
\end{enumerate}

A clear advantage of these methods, as compared to prescribed
synchronization methods, is that they have no sequential scheduling overhead,
since all tasks can be scheduled in parallel, virtually at any point in time. 

The cost of storing synchronization objects (spatial overhead) differentiates
both methods. 
It is $O(n^2)$ for Method 1 (one tag per dependence), and $O(n)$ for Method 2
(one tag per task). 
A subtlety here is that while only completed tasks perform a
\verb|put| in the tag table, all the other tasks may have performed at least one
\verb|get|, which needs to be tracked by the runtime. 
However, Method 1 has an advantage in terms of garbage-collection of its tags,
since a tag can be disposed of as soon as its unique \verb|get| has executed. 
The SWARM runtime offers one-use tags, where the disposal of a tag is performed
by the runtime after a \verb|get| was done on the tag. 
In Method 2, without further sequentializing synchronization, successor tasks
don't know whether they are the last task to \verb|get| the tag. 
Hence the tag objects can only be disposed once a post-dominator of the task
graph has started, or when the entire task graph has completed. 

In terms of in-flight task overhead, a straightforward implementation of both
methods consists in starting all the tasks upfront and letting them synchronize
with each other.
To reduce the number of tasks to be managed by the scheduler simultaneously,
tasks should be scheduled by their predecessors. 
The tightest bound on the number of in-flight tasks is obtained when each task
is scheduled by one of its predecessors. 
In this case, the number of in-flight scheduled tasks is exactly the number of
tasks that are ready to run. 
The main obstacle to achieving this appears again with tasks that have more than
one predecessor, in which case one of the predecessors must be chosen to
schedule the successor task. 
This cannot be performed dynamically, using tags only, without introducing
sequentializing synchronizations. 

Methods for statically electing a task within a set of tasks are available in
the context of automatic parallelization using the polyhedral model. 
For any given successor task, the method consists in considering a total order
among the set of predecessor tasks, and defining the task that minimizes the
order \cite{feautrier88parametric, isl} as the one that schedules the successor
task.
Unfortunately, except for simple cases, the computation of such a minimum
doesn't scale well with the number of nested loops in the program,
leading to potentially intractable execution times of the automatic
parallelization tool. 
Additionally, this solution is specific to the polyhedral model, and in this
section we are discussing synchronization models in the general context of
automatic generation of EDT codes. 

Hence, here we consider the straightforward solution as the only generally viable
option, with a $O(n)$ in-flight task overhead. 
Method 2, proposed by \cite{vasilache13tale}, is superior to Method 1 across the
board, except for its garbage collection overhead. 

\subsubsection{Counted dependences} \label{ssec:counted}
Counted dependences have similarities with both prescribed synchronizations and
tags.
Like prescribed synchronization, they require a task to initialize them. 
A counted dependence naturally represents the number of unsatisfied input
dependences of a successor task.
It is decremented by each predecessor of the task, at completion.
Without further synchronization, counted dependences have an $O(n^2)$
sequential overhead -- like prescribed synchronizations -- if the input
dependences need to be enumerated, or $O(n.d)$ if there exists an analytic
function that computes them in time $d$.

Such a function can be generated in the case of polyhedral code generation. 
We show in section \ref{sec:deps-codegen} that task dependences can be
represented with a polyhedron, which scans the predecessors (resp. successors)
of a task as a function of the task's own runtime parameters. 
We use polyhedral counting techniques to compute the number of predecessors to a
task, either by evaluating the enumerator of the
polyhedron~\cite{clauss97deriving,1275134}, or by scanning the polyhedron as a
loop and incrementing the count by one for each iteration.
The best choice of a counting loop versus an enumerator depends upon the shape of the
polyhedron. 
Complex shapes result in complex enumerators, which can be costlier to
evaluate in practice than with a counting loop, especially if the count is low.

The fact that the program starts with $n$ tasks to schedule implies an in-flight
task overhead of $O(n)$.
Only one counted dependence is required for each task, giving a spatial overhead
and an in-flight overhead of $O(n)$. 
Garbage collection of the counted dependence associated with each task can be
performed as soon as the task starts. 

The set of useful in-flight scheduled tasks should be the ones that are
ready to run, i.e., the ones whose input dependences are satisfied, plus the
ones that are already running. 
Let $r$ be the maximum number of such tasks in any execution of the task
graph.
Having less than $r$ in-flight tasks would reduce parallelism and is hence not
desirable.
An ideal task graph runtime scheme would have $O(r)$ in-flight task overhead
and, accordingly, an $O(r)$ spatial overhead.

\subsubsection{Autodecs} \label{ssec:autodec}
Sequential overhead results from the inability to determine, for a given
successor task with multiple predecessors, a unique predecessor that can set up
the successor task (let us call such task the successor's creator).

As we saw in previous sections, we assume that there is no general, viable way
to resolve this statically. 
We propose a dynamic resolution based on counted dependences, which does not
introduce sequentializing dependences. 
Again, this is different from dynamic dependence discovery as performed by some
runtimes, since the dependences here are defined by the compiler. 
In our proposed dynamic resolution, the first task to be able to decrement a
successor task's counter becomes its unique creator. 
We call such decrement with automatic creation an \verb|autodec| operation. 

Creation of a unique counted dependence -- and hence a unique successor task --
can be ensured for instance using an atomic operation, which deals with the
presumably rare case when two predecessors would complete at the exact same
time.

As a result, only tasks that don't have a predecessor need to be scheduled by
the master task (no sequential start-up overhead). 
The tasks that have predecessors are scheduled upon the first completion of one
of their predecessors, resulting in an $O(o.r)$ in-flight dependence overhead. 
Tasks are only scheduled when all their dependences are satisfied, resulting in
an $O(r)$ in-flight task overhead. 

An $O(r)$ spatial overhead can be obtained by storing the counted dependences in
a map (for instance a hash map), at the price of more complex synchronization
mechanisms.

Consider the case where the set of tasks without predecessors is unknown
statically. 
Since we know the set of predecessors for each task, one solution would be to
identify this set by scanning all the tasks and collecting the ones with zero
predecessors. 
Unfortunately, this would entail a worst-case sequential start-up overhead of
$O(n)$, which can be dramatically optimized in the case of the polyhedral model,
as presented in section \ref{ssec:codegen-autodec}. 

To avoid this need, we introduce a \verb|preschedule| operation, in which a
counted dependence is atomically initialized -- as in autodecs -- but not
decremented.
The fact that the same mechanism is used by autodec and preschedule operations
guarantees that no counted dependence will be created more than once, and that
no task will be executed more than once. 
Hence, the order in which the master task preschedules tasks and tasks
auto-decrement their successors does not matter, and preschedule operations can
execute concurrently with the tasks, resulting in a $O(1)$ sequential start-up
overhead. 

\paragraph{Porting autodec principles to the tag-based model:}
A similar synchronization combined with task initialization could be implemented
on top of Tags Method 1, using an ``auto-put'' operation, through which
the first predecessor to a task also sets up the task. 
Unfortunately, this method would still suffer from a higher spatial overhead
($O(r^2)$), since one tag is associated with each dependence. 

It is clear that counting could be used in Tag Method 2 \cite{vasilache13tale}
to reduce its garbage collection overhead to $O(r)$. 
However, we do not see a way around the $O(n)$ overhead for spatial occupancy
and number of in-flight tasks. 



\section{Scalable task dependence generation} \label{sec:scalable}


Automatic extraction of task parallelism is an attractive proposition. 
Unfortunately, existing techniques based on the polyhedral model
weaken this proposition, because their practicality is limited by
their poor algorithmic tractability.

For the sake of simplicity, in this section we assume that each tile defines a
task, and we use both words interchangeably.
In practice, a task is defined either as a tile or as a set of tiles. 
Also, a useful guideline is that no synchronization should happen inside a 
task, which enables the scheduler to prevent any active wait. 

The base technique used by \cite{Baskaran2009, dathathri14dataflow, kong14taco}
to compute tiles and tile dependences is as follows.
The authors form dependence relationships among pairs of tiled references.
The dependence domain is expressed in the Cartesian product of the tiled
iteration
domains of the source and destination (polyhedral) statements
The task dependences are obtained by projecting out the intra-tile dimensions in
both source and destination iteration spaces. 
In \cite{kong14taco}, the transformations are actually expressed in terms of a
transformation from the iteration domain to a multi-dimensional time range
called the schedule. 
Useful schedules being bijective functions, descriptions
based on the domain and the schedule are equivalent in practice. 
Here, we choose to use the domain-based description because it is simpler. 

Unfortunately, the base technique does not scale well because it relies on the
projection of a high-dimensional polyhedron (or of integer-valued points in the
polyhedron).
Projection is know to scale poorly with the number of dimensions of the
source polyhedron.
This is true even when the rational relaxation of the source polyhedron
is considered, a valid and slightly conservative approximation in the case of
dependences. 

Here, we present a technical solution which does not require the computation of
a high-dimensional dependence domain, and also does not rely on projections. 
Our technique assumes that iteration space tiling partitions computations
into parallelotopes. 
In current polyhedral parallelization, hyperplanes that define the shape
of the parallelotopes are defined by scheduling hyperplanes. 
Together, they form a schedule, which defines a transformation of the domain. 
Tiling is then performed along these hyperplanes. 
Since we are eliding the schedule in this description, parallelotope tiling
hence corresponds to applying the transformation defined by the scheduling
hyperplanes to the iteration domain, followed by orthogonal tiling. 
Hence, without loss of generality and for the sake of clarity, we are describing
our method assuming orthogonal tiling, where the tiling hyperplanes are defined
by canonical vectors of the iteration space. 

The main idea is to start with a {\em pre-tiling} dependence (i.e., among
non-tiled iterations), and to derive the inter-tile dependences by expressing
the tile iteration spaces using a linear compression of the pre-tiling iteration
spaces.

Sets of integer-valued points are represented in the polyhedral model by
a rational relaxation, which can be represented compactly as the integer points
of a (rational) polyhedron.
We first explain our technique on a polyhedron $D$, for which we consider
a tiling transformation defined by a matrix $G$. 
We show how to precisely define the set of tile indices that correspond to
tiles that contain integer points in $D$.
More specifically, let the integer diagonal matrix with positive diagonal
elements $G \in \Z^{n\times n}$ represent the orthogonal tiling transformation
 being applied to the space of index $I\in \Z^n$ in which $D$ is immersed.
 
The relationship between an iteration $I$ and the inter-tile $T\in\Z^n$ and
intra-tile $X\in \Z^n$ dimensions obtained by tiling $I$ according to $G$ is:
\begin{equation} \label{eq:compression}
I = GT + X
\end{equation}
\begin{equation} \label{eq:modulo}
0 \leq X \leq diag(G)-\one
\end{equation}
, where $diag(G)$ is the vector made of the diagonal elements of $G$, and
$\one$ is a $n$-vector of coefficients 1. 

$G$ being invertible, (\ref{eq:compression}) can also be written as:
\begin{equation} \label{eq:compression2}
T = G^{-1}I - G^{-1}X,
\end{equation}
And (\ref{eq:modulo}) can be written as:
\[
0 \leq G^{-1}X \leq \frac{diag(G)-\one}{diag(G)}
\]
where elementwise division is used. 

Let $U$ be defined as: 
\begin{equation} \label{eq:U}
\{Y\in \Q^n: Y = -G^{-1}X, 0 \leq X \leq diag(G)-\one \}
\end{equation}

From Equation \ref{eq:compression2}, any $T$ corresponding to an integer point
$I$ in $D$ is defined by: 
\begin{equation}
T = G^{-1}I + Y, Y \in U
\end{equation}
Hence, the set of values of $T$ corresponding to integer points in $D$ is given
by
\begin{equation} \label{eq:comp-n-plus}
T \in image(D, G^{-1}) \oplus U
\end{equation}
where $\oplus$ represents the polyhedral direct sum operator. 
This set is {\em exact} for any given $D$ and tiling $G$ of $D$'s space. 

We can apply the same method for a dependence $\Delta(I_s, I_t)$ linking the
iteration spaces $I_s$ and $I_t$ of a source statement $s$ and a target
statememnt $t$. 
Let us consider tiling $G_s$ for $s$ and tiling $G_t$ for $t$. 
The corresponding inter-tile dependence $\Delta_T$ is the set of inter-tile
indices $T_s$ and $T_t$ that correspond to an integer point $(I_s, I_t)$ in
$\Delta$. 

We consider the combined compression transformation $G_{s,t}$ which applies $G_s$ to
the $I_s$ space and $G_t$ to the $I_t$ space:
\[
G_{s,t} = \matrice{cc}{G_s & 0 \\ 0 & G_t}
\]
Let $X_s$ and $X_t$ be the intra-tile dimensions defined by $G_{s,t}$.
We get a definition of $U$ in the combined source-target space as in Equation
\ref{eq:U}:
\begin{equation} \label{eq:sum1}
-G^{-1}(X_s, X_t)^T \in U_{s,t}
\end{equation}

Here too, since $G_{s,t}$ is invertible, we define $P = image(\Delta,
G_{s,t}^{-1})$ and we have:
\begin{equation} \label{eq:sum2}
\Delta_T = P \oplus U_{s,t}
\end{equation}

We can hence define an exact inter-tile dependence relationship without
resorting to forming high-dimensional polyhedra and, more importantly, without
having to project any high-dimensional polyhedron. 
The only operations we have used are a linear, invertible compression, and a
polyhedral direct sum. 

While already much more scalable (and we will validate this later on), we could
still look for an even more scalable solution. 
In particular, the direct sum of a polyhedron with a hyper-rectangle is that it
will results in a polyhedron with many vertices, which could reduce the
scalability of further operations.
This can be addressed using a cheap, constraints-oriented way of computing a
slight over-approximation of this particular type of direct sums, presented in
the next section.

\subsection{Preventing vertex explosion}
The following technique for reducing vertices in the task dependence polyhedron
relies on slightly shifting the constraints of $P$ outwards, until the
modified $P$ contains all the points of $P\oplus U_{s,t}$. 
We call this operation an {\em inflation} of $P$ w.r.t $U_{s,t}$.

As stated above, the $U$ polyhedron defined in (\ref{eq:U}) can be written (in
the $T$ space) as:
\[
-\frac{g_i-1}{g_i} \leq T_i \leq 0
\]
$U$ is a hyper-rectangle. 
Its vertices are defined by vector $(g')$, where 
\[
g_i' = \system{c}{
  0 \mbox{ or } \\
  -\frac{g_i-1}{g_i}
}, i \in [1, n]
\]

Consider a constraint of $P$, written as $aT+b \geq 0$, where $b$ may contain
parametric expressions. 
We are looking for an offset $c$ such that $aT+b + c \geq 0$ contains all the
points of $P \oplus U$. 

In other words, 
\[
a(T+(g')) + b + c \geq 0 \Leftrightarrow aT + a(g') + b + c \geq 0
\]
This relationship is respected whenever $c \geq -a(g')$. 
The maximum value for the right-hand side occurs when $g_i'=\frac{g_i-1}{g_i}$
whenever $a_i$ is positive. 
Hence the maximum required value for $c$ is:
\[
c_{max}(a) = \sum_i a_i.g_{a,i} \mbox{, where }
g_{a,i} = \system{c}{
\frac{g_i-1}{g_i} \mbox{ if } a_i >0\\
0 \mbox{ otherwise}
}
\]

The inflated polyhedron is then defined by replacing the constant offset $b$
with $b + c_{max}(a)$ for every constraint of $P$. 
Of course, the $U$ considered for tile dependences is $U_{s,t}$ and the $G$ is
$G_{s,t}$.
Since the inflated task dependence polyhedron is obtained only by shifting
constraints of $P$, it has the exact same combinatorial structure, i.e., we
haven't incresased the number of vertices or constraints through inflation.

\section{Generating dependence code}
\label{sec:deps-codegen}


\subsection{Prescribed Dependences}

In the case of prescribed dependences, the tasks and all their dependences
must be declared before starting the program execution. A task is then executed
by the runtime layer as soon as all its dependences are satisfied.  Polyhedral
compilers maintain during the whole compilation the exact set of iterations and
dependences as polyhedra, which is translated into the required
synchronization APIs.

\begin{figure}
\begin{center}
    \includegraphics[width=1\columnwidth]{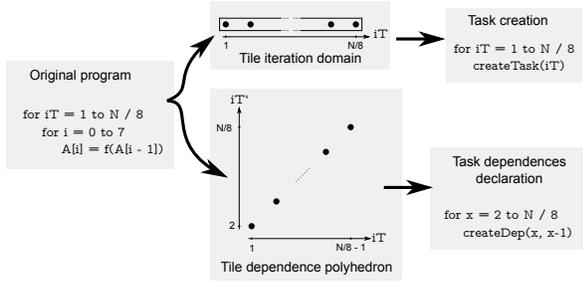}
\end{center}
\caption{For prescribed dependences, the task creation loop is generated from
the tile iteration domain. Task dependences are also declared using the
tile dependence polyhedra. The control code generated to handle cases
where \texttt{N} is not divisible by 8 is not represented to simplify the
notation.}\label{fig:codegen_prescribed}
\end{figure}

Task-based parallelization in the polyhedral model relies on forming
tasks from tiled loop nest. Here again, for simplicity we assume a task is
associated with each instance of a tile, i.e., each value of the inter-tile
iterations corresponds to a task instance. 
Tasks are generated as functions (or their equivalent in the targeted runtime),
whose parameters include their inter-tile coordinates. 
Let the ``tile iteration domain'' of the tiled statements be the set of valid
inter-tile iterations, which corresponds to the set of non-empty tasks. 
The tile iteration domain can be formed by using the
compression method of section \ref{sec:scalable} or by projecting the iteration
domain on its inter-tile dimensions. 
As illustrated in the top of Figure~\ref{fig:codegen_prescribed},
the tile iteration domain is then assigned to a task initialization primitive.
A similar tile creation loop nest is created for every
distinct tiled loop nest in the program, which results in the initialization
code for all the program tasks.


Once tasks are known to the runtime, task dependences are declared. 
As described in section \ref{sec:scalable}, dependences are formed as a
polyhedron in the Cartesian product of the tile iteration spaces of the source
and destination tiles.
A dependence polyhedron defines a relationship between the inter-tile
coordinates of its source tasks and the inter-tile coordinates of its
destination tasks.
Since, in the prescribed model, the role of such polyhedron is to declare the
existence of a dependence, in this section we call it the {\em declarative
  dependence polyhedron}.
Hence, as explained in \cite{Baskaran2009}, they can naturally be generated as
loop nests that scan all the (source task, destination task) pairs that are
connected by a dependence.
A function call is generated as the body of these loops, which declares the
existence of the dependence for each such pair, as illustrated in
Figure~\ref{fig:codegen_prescribed}. 
The loop indices are used
as the coordinates of the task at the origin and at the destination of the
dependence.  As shown in our example, the generated loop nest benefits from any
of the loop optimizations applied during code generation, including their
simplification.

\subsection{Tags}
It is possible to generate code for Tag Methods 1 and 2 from the
declarative form defined above. 

In Method 1, each task first gets a tag from each of their predecessors,
performs computations, and puts a tag for each of their successors. 
The get and put loops can be derived directly from the declarative dependence
relationship, by mapping the destination inter-tile loops of the dependence
polyhedron to the parameters of the task. 
The task performing the \verb|get|s acts as the destination of the dependence
relationship.
A loop that scans all the coordinate of the predecessors as a function of the
inter-tile parameters of the task is generated from the resulting polyhedron,
executing the \verb|get|s.
Symmetrically, the iteration domain of the \verb|put| loop is obtained by
mapping the source inter-tile dimensions of the declarative dependence to the
inter-tile parameters of the task. 

Method 2 is simpler in that each task runs a single \verb|put| call, with its
own inter-tile parameters as parameters to the \verb|put|. The \verb|get|s are
obtained in the same way as for Method 1. 

\begin{figure}
\begin{center}
    \includegraphics[width=1\columnwidth]{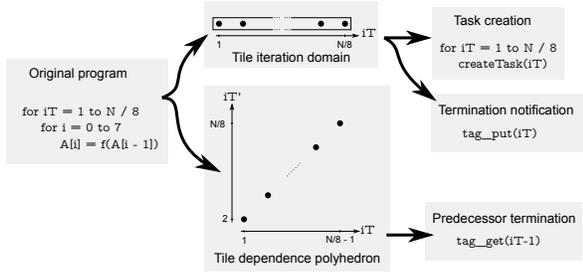}
\end{center}
\caption{For tags (Method 2), each task issues a \texttt{put} operation for
  itself and every task waits for its predecessors using a tag \texttt{get}
  operation.
The control code generated to handle cases where \texttt{N} is not divisible by
8 is not represented to simplify the notation.}\label{fig:codegen_tags}
\end{figure}

The process is illustrated in
Figure~\ref{fig:codegen_tags}, where the optimizations performed during code
generation simplify the complex loop nest into a single
\texttt{get} statement parameterized by the task coordinates \texttt{iT}.

\subsection{Autodecs}\label{ssec:codegen-autodec}

Autodecs use counted dependences and atomic task initialization to enable tight
dynamic task scheduling.
With autodecs, only tasks with no predecessor need to be created by the master
EDT. 
When a task ends, it iterates over all its successors in order to decrement its
number unsatisfied dependences. 
Such a loop is generated precisely like the \verb|put| loop in Tag Method 1,
except that the function called is \verb|autodec| instead of \verb|put|.

The first task which decrements the incoming dependence counter of any of its
successors also initializes the successor's counted dependence. 
To do so, it needs to compute the number of predecessors of said successor task. 
In order to implement this, a predecessor count function is made available
specifically for autodecs by the compiler. 
This function takes the successor task's inter-tile coordinates and returns the
number of its predecessors.
The number of predecessors is defined by the number of integer-valued points 
in the dependence polyhedron, as a function of the successor task's inter-tile
coordinates.

There are two possible ways of generating such a function, and both can be
defined from the get loop from Tag Method 1. 
One way is as a loop, by turning the \verb|get| calls into increments of a
counter. The returned value is the number of iterations in the loop, i.e., the
number of predecessors to the task. 
Another way consists in computing the enumerator of the \verb|get| loop nest,
i.e., an analytic function, which returns the number of integer-valued points in
the polyhedron representing the \verb|get| loop, as a function of the inter-tile
parameters of the task. 

Heuristics to determine which form is best are essentially based on the shape of
the polyhedron representing the \verb|get| loop and an estimate of the number of
iterations.
Enumerators are not sensitive to the number of iterations, but very much to the
shape of the dependence polyhedron, while direct iteration counting is
insensitive to shape but undesirable when the number of predecessors is high.

\begin{figure}
\begin{center}
    \includegraphics[width=1\columnwidth]{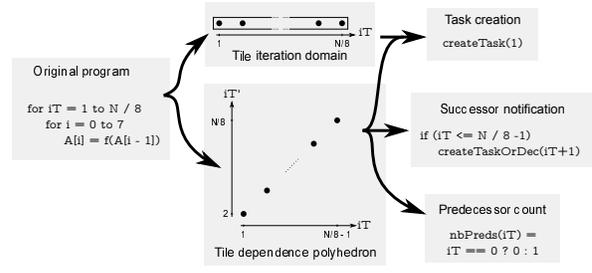}
\end{center}
\caption{With autodecs, only tasks without predecessors are explicitly
initialized. Every task computes its number of predecessors. 
At completion time, tasks decrement the input dependence counter of their
successors, and may initialize them. 
The control code generated to handle cases where \texttt{N} is not divisible by 8
is not represented to simplify the notation.}\label{fig:codegen_autodecs}
\end{figure}

The resulting expression defines the exact number of predecessors for a task as
a function of the task coordinates and problem parameters.  The generated code
is illustrated in Figure~\ref{fig:codegen_autodecs}.

As opposed to the previous methods, with autodecs, the tasks are created
by one of their predecessors. A loop scanning the tasks without predecessors
has to be created for execution by the master task. 

To determine the set of tasks without predecessor, we project the dependence polyhedra
on their destination dimensions. 
The projected polyhedra represent the coordinates of all the tasks with a
predecessor. 
The projected polyhedra are then subtracted from the target statement iteration
domain, which results in the set of tasks without predecessors.



\section{Experiments} \label{sec:exp}
In order to validate our findings, we compare the computation time of our
polyhedral dependence computation method with the current state-of-the-art
methods \cite{Baskaran2009, kong14taco} in section \ref{ssec:exp-compression}.
Then, in section \ref{ssec:exp-overhead} we explore
the question of the significance of the worst-case complexity analysis we
presented in section \ref{sec:deps-comparison} by comparing concrete values for
overheads with high worst-case complexity.

``Machine A'' is a 12-core, dual-hyperthreaded Xeon
E5-2620 running at 2.00GHz with 32Gb RAM running Linux Ubuntu 14.04.
Our version of the parallelizing compiler produces source code, which we compile
with GCC 4.8.4. 
``Machine B'' is a 32-core, dual-hyperthreaded Xeon E5-4620 running at 2.20Ghz
with 128Gb RAM running Ubuntu Linux 14.04. 

\subsection{Compile-time dependence computation scalability}
\label{ssec:exp-compression}
While it is hardly debatable that performing a linear compressions of
low-dimensional polyhedra is (much) less computationally expensive than
projecting roughly half the dimensions of a high-dimensional polyhedron, a
few well-chosen experiments could help evaluate the importance of the problem.

\begin{figure*}
\begin{center}
\includegraphics[width=1.5\columnwidth,height=4cm]{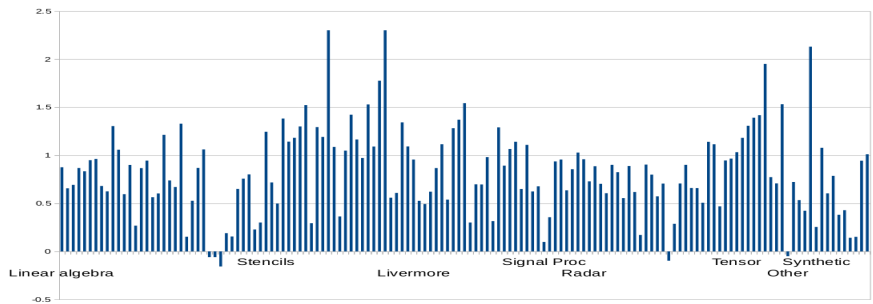}
\caption{$log_{10}$ of Speedups of the compression method over the projection method}
\label{fig:dep-speedups}
\end{center}
\end{figure*}

In order to perform a meaningful comparison, we enforced the same behavior of the
polyhedral optimizations upstream (such as affine scheduling and tiling), by
running the tool with default options. 
We also turned off the removal of transitive dependences, so as to leave
discussions about trade-offs between compilation time and precision of the
dependences out of the scope of this paper. 
Transitive dependence removal hardly decreases the dimensionality of the
problem and increases the number of dependence polyhedra. 
We instrument the code in order to measure dependence computation time only over
143 benchmarks which include linear algebra, radar and signal processing codes
(including FFT-based), stencil computations, sparse tensor codes, an
implementation of the Livermore benchmarks \cite{livermore-loops}, and a handful
of synthetic codes.
The speedups on Machine A are reported on a logarithmic scale in Figure
\ref{fig:dep-speedups}. 

The high points are as follows.
\begin{itemize}
\item Two benchmarks exceeded the 3-minute timeout in the naive method. 
  While this is a global compilation time timeout, the bottleneck there was
  clearly the computation of the task dependences. 
  These two benchmarks were taken out of the measurements below. 
\item The average speedup is 10.5X, the maximum (excluding timeouts) is
  135X. In order to improve the readability of Figure \ref{fig:dep-speedups}, we
  arbitrarily capped the timed-out compilations to a 200X speedup. 
  These numbers imply great practical compilation time speedups, considering that
  these operations are typically
  the computational bottleneck in the parallelization process. These speedups are
  relatively low, considering the combinatorial nature of the problem.
  This is explained by the fact that we are only compiling the code with one level
  of tiling. 
  If tiling (or any other strip-mining-based transformation) were used to target
  more than one level of processing or memory, the base number of iteration dimensions
  would increase significantly, and the gap between our compression method and
  the projective method would increase dramatically. 
\item There are a few cases where the projection method is slightly faster than
  the compression method.
  We looked at these cases, and the simple explanation there is that the
  projection is very efficient for these iteration domains.
\item Some dependences are computed even for codes that are usually seen as
  ``embarrassingly parallel.'' There are two reasons for this. 
  First, code is often partitioned in different ways than with loop
  parallelism, in order to create more tasks and increase load balancing. 
  For instance, in matrix multiplication, a task is created for each tile, i.e.,
  each iteration of the three outer loops (including the reduction loop). 
  Also, we have turned off optimizations that simplify the dependences based on
  known parallelism information, hence dependences are sometimes built, only to
  realize later that they are empty.
\end{itemize}

\subsection{Worst-case overheads}
\label{ssec:exp-overhead}
The problem of creating meaningful comparisons among codes generated for
different synchronization models is somewhat difficult for a few reasons. 

First, with OCR we can use prescription and autodecs, while with SWARM we can
use tags and autodecs. 
Comparisons among specific runtimes being out of topic for this paper, we decide
to only perform execution time comparisons within each runtime. 
We compare execution times obtained with prescription and autodecs in OCR, and
compare those obtained with tags and autodecs in SWARM. 

Similarly, a large space of schedules and tile sizes can be explored in the
process of optimizing EDT codes. 
While they have a great impact on performance and are an interesting topic in
and of themselves, they are not quite relevant to the problem we are addressing
in this paper. 
Hence, instead of comparing best execution times among all the possible compiler
optimizations we could apply, we chose to pick a particular parallelization
choice (the one obtained using the default settings of the compiler), and
compare execution times obtained across synchronization models. 

The goal of these comparisons is to evaluate the relevance of the worst-case
overhead figures we derived in section \ref{sec:deps-comparison}. 
With Exascale at the horizon, we are easily convinced that they will eventually
do, but estimating their impact on current machines is informative of how soon
we should start worrying about them.

We do not intend to be exhaustive here but just understand trends, and hence we did
not reimplement Tag Method 2 \cite{vasilache13tale}, which was neither the optimal
nor presenting the most serious overhead behaviors.
 
We ran a sample set of benchmarks both in prescribed mode and autodec mode using
OCR and compared their execution times on machine B. 
We observed speedups in a majority of the benchmarks (up to 27X for a
fixed-point Givens QR code), but roughly a third of them have slowdowns, up to
5X for a synthetic benchmark. 
Quite systematically, the benchmarks for which the autodec version is slower
have short execution times, mostly below 0.1 s, suggesting that they
correspond to a small number of tasks.   

We also ran a sample set of benchmarks both using Tag Method 1 and autodecs in
SWARM. 
Speedups are more salient there, as autodec-based versions are up to 75X faster
(for the \verb|trisolv| benchmark), and one slowdown is observed (10X for
\verb|covcol|).
This shows that both the synchronization model and the way it is employed should
not be overlooked, as they have a major impact on the generated program. 
Also, three benchmarks did not finish in the tag-based implementation because
they ran out of memory, while they do not run out of memory using autodecs.
This shows that the $O(n^2)$ spatial cost can be {\em already limiting} on a
32-core machine. 


\section{Related Work} \label{sec:related}

The most directly related work targets task-based parallelism from a polyhedral
program representation. 
In particular, Baskaran et al proposed a task-based strategy for
multicore processors using the polyhedral model~\cite{Baskaran2009}. The
strategy is not intended for large scale systems and requires the full set of
tasks dependences to be expressed before starting the execution. Our approach
has lower memory and computational requirements and is then much more scalable.
Moreover, even though tile dependences are considered, they are obtained using
polyhedral projection, which is computationally costly. A similar approach is
considered more recently by Kong et al~\cite{kong14taco}, whose focus is on the
generation of OpenStream code. The same projection-based method is also used in
the distributed dataflow work of Dathathri et al\cite{dathathri14dataflow}.
In our method, the tile dependences are deduced from the original program
representation, and without requiring any intractable polyhedral projection.

The various proposed solutions focus on different aspects of the
polyhedral representation and are often complementary.

An important point about the compression technique is that it addresses one
important tractability issue in performing computations on polyhedra during
compilation.
Tractability is a core problem in polyhedral compilation. 
It cannot reasonably be ignored in a production compiler. 
Hence, much work has been performed to improve tractability of polyhedral
compilation in the literature, often at the price of approximations or by
introducing extra constraints on the program representation.
Several techniques restrict the set of constraints allowed to define polyhedra.
Several variants of the same techniques exist, each one restricting differently
the form of the constraints that can be handled. For instance, Difference
Bound Matrices (DBM) only allows constraints in the form $x_i - x_j \le k, x_i
\ge 0, x_j \ge 0$~\cite{Shaham00,Mine01DBM}. Other representations allow more
complex constraints such as Unit Two Variables Per Inequality
(UTVPI)~\cite{Bagnara09,Mine06} or Two Variables Per Inequality
(TVPI)~\cite{Simon03TVPI,upadrasta12-two-vars-per-ineq-impact,upadrasta13-popl}
for instance. The general idea is to restrict the form of the constraints in
order to use specialized algorithms to handle the polyhedra, usually with
strong worst-case complexity guarantees. In a different direction, Mehta and
Yew recently proposed to over-approximate a sequence of statements as a single
element called O-molecule~\cite{Mehta15}. Their approach reduces the number
of statements considered in a program, which drastically improves the
complexity of several polyhedral operations performed during compilation. A
similar solution was proposed by Feautrier~\cite{Fea04}, and can also be
perceived in the work of Kong et al~\cite{kong14taco}. All the cited
improvements are independent from our work and can be combined with the
dependence analysis based on tiles presented in this paper.

An alternative, scalable approach to computing tile dependences requires the
programmer to express their program in terms of computation (and data) tiles, as
in \cite{parsec}.

The second contribution of this paper improves the scalability of the runtime in
charge of scheduling the tasks. This is in a context where the runtime is
given all dependences by the programmer, as opposed to runtimes that discover
dependences as a function of data regions commonly accessed by tasks
(as in \cite{legion, starss}). 

Our proposed improvement does not rely on new
language constructs and can be achieved automatically, without involving the
programmer. Moreover, the task dependence management we propose is not
specifically related to any runtime system, although some of them are better
candidates for an integration. We successfully implemented our optimization for
two different runtimes: SWARM~\cite{swarm}, and OCR~\cite{ocr-spec-100}. 
Furthermore, nothing would prevent the implementation of our optimizations on any
 runtime that enables the composition of programs as a graph of tasks.
On the other hand, in some executions models Cilk~\cite{Blumofe:ACM:1995} or
X10~\cite{x10}, the task graph is supported by a tree in which tasks can synchronize
with their direct or (respectively) transitive children, which is less general
than the model we considered.
Such models usually provide termination guarantees in exchange for reduced
generality of the task graph model. The polyhedral model provides similar
guarantees without specifically imposing tasks trees, although its application
domain is more limited than what can be written by hand using tree-supported
languages. 
OpenStream~\cite{PCo13} also provides a less restricted task model, although it
seems geared towards much finer synchronization granularity based on streams of
data.

\section{Conclusion} \label{sec:conclusion}
We presented a truly scalable solution for the generation of event-driven task
(EDT) graphs from programs in polyhedral representation by a compiler.  We
investigated tractability issues in both the compilation time and the execution
time of such generated programs, and offered two main contributions.

First, we explored the use of three different synchronization models available
in current EDT runtimes. 
We evaluated their overheads in terms of space, in-flight task
and dependence management, and garbage collection. We found out a way of using
a slight extension of one of the synchronization models to reach
near-optimal overheads across the board. 

Second, we contributed a method to dramatically reduce the computation time of
the costliest operation required to generate EDT codes in a polyhedral compiler:
the generation of inter-task dependences. 
We also discussed how to generate code for the three synchronization models from
their polyhedral representation. 

Both aspects unlock limitations of EDT code generation for polyhedral
compilers. 
We also believe that our comparative study on synchronization models is useful
to anyone who would want to implement an automatic code generation framework
based on the ones we considered.

These methods were fully implemented in the R-Stream
compiler~\cite{rstream2011}, using the OCR~\cite{ocr-spec-100} and SWARM~\cite{swarm}
runtimes.
More optimizations related to inter-task dependences have an impact on
performance. 
Some of them were addressed in the literature, but there are more to be done. 

Our polyhedral tile dependence computation method supports most practical
tilings out-of-the-box, including diamond tiling.
Nevertheless, extensions to more exotic -- but useful -- tilings, such as hexagonal
tiling or overlapped tiling, would be of interest as well.

\section{Acknowledgements} \label{sec:ack}
This research was developed with funding from the Defense Advanced Research
Projects Agency (DARPA).
The views and conclusions contained in this document are those of the authors
and should not be interpreted as representing the official policies, either
expressly or implied, of DARPA or the U.S. Government.


\bibliographystyle{plain}
\bibliography{pca}
\end{document}